# Effect of magnetic perturbations on turbulence-flow dynamics at the L-H transition on DIII-D

D. M. Kriete[1], G. R. McKee[1], L. Schmitz[2], D. R. Smith[1], Z. Yan[1], L. A. Morton[1], R. J. Fonck[1]

[1]Department of Engineering Physics, University of Wisconsin–Madison, Madison, WI 53706, USA
[2]Department of Physics, University of California Los Angeles, Los Angeles, CA 90095, USA

E-mail: kriete@wisc.edu

**ABSTRACT**

Detailed 2D turbulence measurements from the DIII-D tokamak provide an explanation for how resonant magnetic perturbations (RMPs) raise the L-H power threshold $P_{LH}$ [P. Gohil *et al.*, Nucl. Fusion **51**, 103020 (2011)] in ITER-relevant, low rotation, ITER-similar-shape plasmas with favorable ion $\nabla B$ direction. RMPs simultaneously raise the turbulence decorrelation rate $\Delta\omega_D$ and reduce the flow shear rate $\omega_{shear}$ in the stationary L-mode state preceding the L-H transition, thereby disrupting the turbulence shear suppression mechanism. RMPs also reduce the Reynolds stress drive for poloidal flow, contributing to the reduction of $\omega_{shear}$. On the ~100 μs timescale of the L-H transition, RMPs reduce Reynolds-stress-driven energy transfer from turbulence to flows by an order of magnitude, challenging the energy depletion theory for the L-H trigger mechanism. In contrast, non-resonant magnetic perturbations, which do not significantly affect $P_{LH}$, do not affect $\Delta\omega_D$ and only slightly reduce $\omega_{shear}$ and Reynolds-stress-driven energy transfer.

## I. INTRODUCTION

Next-step tokamaks that aim to produce burning plasmas envision operating in the high-confinement mode (H-mode)[1] due to its high achievable density and temperature. Steep gradients arising from the H-mode edge transport barrier can trigger explosive magnetohydrodynamics instabilities called edge-localized-modes (ELMs).[2] ELMs rapidly expel heat and particles from the plasma and in a reactor-scale tokamak will cause unacceptably high erosion of the divertor. Therefore, ELMs must be controlled or avoided outright. A leading technique for ELM control is the application of resonant magnetic perturbations (RMPs) to the plasma edge, which has been successfully demonstrated to suppress ELMs on several tokamaks, including DIII-D,[3] KSTAR,[4] EAST,[5] and ASDEX-Upgrade.[6,7]

In a reactor-scale tokamak, RMPs will need to be applied before the first large ELM. Applying RMPs in the ~100 ms ELM-free period following the L-H transition may be possible but could be operationally challenging due to the finite response time of the RMP coils. RMPs may therefore need to be applied before the L-H transition. However, applying RMPs to L-mode plasmas raises the L-H power threshold $P_{LH}$, inhibiting H-mode access.[8–14] This is particularly a concern for the pre-fusion power operation (PFPO-1) phase of ITER since available heating power will only marginally exceed the value of $P_{LH}$ predicted by the empirical Martin scaling, which does not account for RMP effects.[15,16] In contrast, non-resonant magnetic perturbations (NRMPs), which do not suppress ELMs, have little effect on $P_{LH}$.[9,17] Even in the absence of applied RMPs, small intrinsic magnetic perturbations arising from error fields can affect $P_{LH}$.[13] Physically understanding how RMPs alter the dynamics of the L-H transition may help predict the $P_{LH}$ increase for future tokamaks, aid in developing techniques that mitigate this effect, and more generally inform the construction of a physics-based predictive model for $P_{LH}$.

The L-H transition is a complex phenomenon wherein turbulence near the edge of magnetically confined plasmas is rapidly suppressed, causing the plasma to bifurcate to a state with markedly improved confinement. While the L-H transition is not yet understood at the level of detail required to predict $P_{LH}$, there *is* broad consensus on the major physics elements involved, namely interactions between plasma profiles, turbulence, and flows. Gradients in the density and temperature profiles provide linear instability drive for turbulence, which in turn produces cross-field transport that acts to relax these gradients. Radially sheared $E \times B$ flow can suppress turbulence,[18] reducing cross-field transport and allowing steeper gradients to form. The L-mode state is characterized by large turbulence-driven cross-field transport and small profile gradients and flow shear. There is significant experimental evidence that turbulence is suppressed in the fully developed H-mode state by mean $E \times B$ shear that is generated by diamagnetic flow.[19,20] The diamagnetic flow is driven self-consistently by the steep pressure gradient at the edge of H-mode plasmas, where turbulence-driven cross-field transport is suppressed.

The mechanism that triggers the L-H transition is more uncertain. There are several theories for the L-H trigger mechanism,[21,22] a leading one of which is the predator-prey model.[23–25,20] This theory posits that the L-H transition occurs when energy in turbulence (the prey) is rapidly transferred via Reynolds stress to zonal flows (the predator), transiently suppressing the turbulence and allowing the edge pressure gradient to begin growing. Turbulence suppression is then maintained by mean $E \times B$ shear. Reynolds-stress-driven energy transfer only becomes significant when the turbulence reaches sufficiently large amplitude. As heating power $P_{in}$ increases, the turbulence amplitude grows until $P_{in}$ reaches a threshold value $P_{LH}$, at which point the energy transfer rate is large enough to deplete the turbulence energy, thereby triggering the L-H transition.

Experimental evidence from several devices supports the predator-prey model.[20] Notably, prior measurements on DIII-D using the same diagnostic and analysis techniques as in this paper showed a large burst of Reynolds-stress-driven shear flow 100 μs before the L-H transition, consistent with the predator-prey model.[26] The same phenomenon was also observed on HL-2A,[27] EAST,[28] and Alcator C-Mod.[29] However, other devices have found evidence against the predator-prey model. Measurements on ASDEX Upgrade showed that Reynolds-stress-driven flow was negligible compared to diamagnetic flow.[30] And work on NSTX found that energy was transferred *out of* zonal flows *into* turbulence right before the L-H transition.[31] The experimental results reported in this paper will further challenge the predator-prey model in the case of L-H transitions with applied RMPs.

Several theories predict that RMPs alter various physics elements of the L-H transition. A fluid model has predicted that RMPs induce a $j \times B$ torque that competes with Reynolds stress drive for zonal flows.[32] Similar modelling work has found that, in the presence of background $E \times B$ shear, RMPs couple zonal flows to Alfvén waves.[33] Gyrokinetic calculations of the residual zonal flow in the presence of radial and parallel magnetic perturbations have shown that zonal flows become collisionlessly damped and decay to zero.[34–36] And a recent 2D MHD model has shown that tangled magnetic fields, which would be present in stochastic regions, strongly modify Reynolds stress phase coherence.[37] All these theories predict that RMPs effectively increase the zonal flow damping rate. According to the predator-prey model, more energy transfer from turbulence to zonal flows would then be needed to trigger the L-H transition. This provides a potential explanation for how RMPs raise $P_{LH}$. Previous measurements also showed that turbulence is significantly increased just inside the pedestal region when RMPs are applied to H-modes, which is consistent with reduced zonal flows and/or increased growth rates.[38]

In this paper, we present experimental evidence that RMPs raise $P_{LH}$ by (i) simultaneously reducing flow shear rates, raising turbulence decorrelation rates, and reducing Reynolds stress in the stationary L-mode state preceding the L-H transition and (ii) disrupting transient Reynolds-stress-driven energy transfer from turbulence to flows during the ~100 μs timescale of the L-H transition. The reduction of Reynolds stress in L-mode contributes to the reduction in flow shear, which together with raised decorrelation rates disrupts the suppression of turbulence by shear flow. To overcome this, more transient turbulence suppression at the L-H transition is needed, requiring more input power to access H-mode. This work is an extension of Ref. 39 and expands on it by including quantitative calculations of Reynolds-stress-driven poloidal flow and nonlinear energy transfer, as well as more details on turbulence characterization.

This paper is structured as follows. Section II discusses the experimental methodology and key diagnostics used to measure fluctuation quantities relevant to L-H transition physics. Section III presents how magnetic perturbations (MPs) alter turbulence and flow characteristics in the stationary L-mode state that precedes the L-H transition. Section IV presents how MPs affect turbulence-flow dynamics on the ~100 μs timescale of the L-H transition. Section V summarizes the major results of this paper and identifies avenues for future research.

## II. EXPERIMENTAL SETUP AND DIAGNOSTIC MEASUREMENTS

To obtain results pertaining to ELM-suppressed H-mode operation on ITER, experiments on DIII-D made detailed turbulence measurements in ITER-relevant plasmas with MPs applied before the L-H transition and the conditions needed to achieve ELM suppression, as described in Ref. 40. The plasmas utilized a lower-single-null, ITER-similar shape (ISS) [Fig. 1(a)] with toroidal field $B_T = 1.95$ T, plasma current $I_p = 1.5$ MA, and safety factor $q_{95} = 3.6$. The ion $\nabla B$ drift direction was towards the X-point, i.e., the direction with lower $P_{LH}$. The line-averaged electron density just before the L-H transition was varied over $\bar{n}_e = 1.5\text{--}5\times10^{19}$ m$^{-3}$, encompassing the density where $P_{LH}$ is minimized, $\bar{n}_e \approx 3\times10^{19}$ m$^{-3}$. All plasmas were heated by balanced neutral beam injection (NBI) to keep rotation low for ITER relevance. Additional electron cyclotron heating (ECH) was injected at mid-radius ($\rho \approx 0.5$) as needed to trigger the L-H transition.

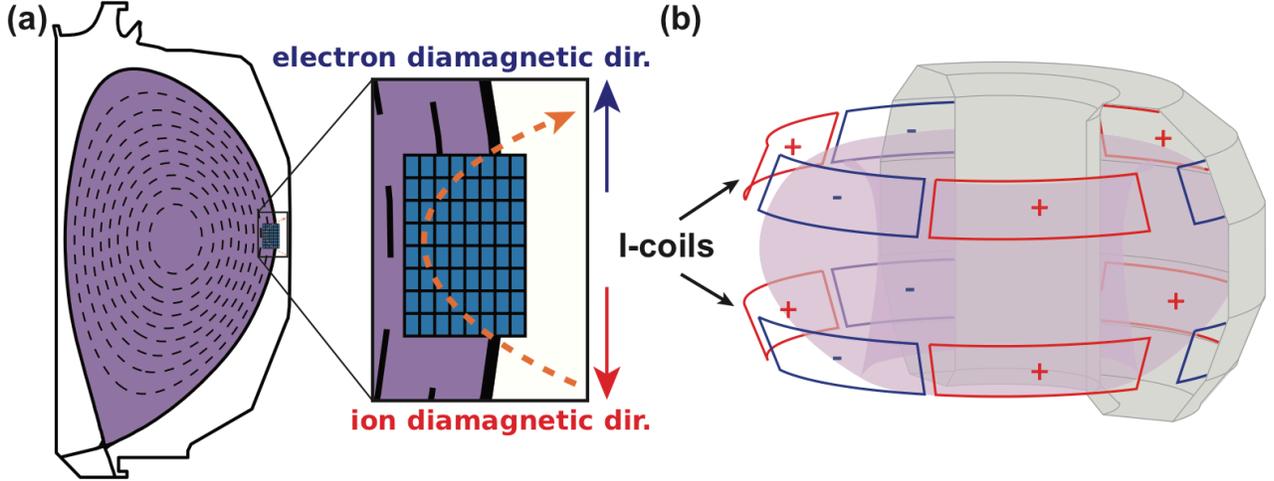

**FIG. 1.** (a) Cross-section of the plasma shape with a zoomed-in view showing the 8×8 grid of beam emission spectroscopy (BES) channels in blue and the ray trajectory for Doppler backscattering (DBS) in orange. (b) Diagram of the in-vessel coil (I-coil) used to apply magnetic perturbations (MPs), shown with the current polarities needed to produce an $n = 3$ resonant magnetic perturbation (RMP).

Magnetic perturbations were applied using the I-coil,[41] which consists of two toroidal rows of six window-frame coils each, i.e., it is a 2×6 coil [Fig. 1(b)]. In contrast, the 3×9 ITER ELM coil[42] will have a wider and more flexible operational space. In this experiment, only MPs with toroidal mode number $n = 3$ were investigated. Previous experiments have also shown that $n = 1$ and $n = 2$ RMPs raise $P_{LH}$.[12,13] Resonant MPs were created by energizing the I-coil in even parity configuration, i.e., with identical current polarity in each upper/lower coil pair. Non-resonant MPs were created using odd parity configuration, where the current polarity is opposite for the upper/lower coils within each pair. There are two possible MP toroidal phase angles: 0° and 60°. Most measurements were made for 0° MPs but some data with 60° MPs was also collected.

We define the resonant $\delta B_r$ component of the applied MP as the pitch-aligned component at the $m/n = 11/3$ surface, which is the closest $n = 3$ rational surface to $\psi_N = 0.95$, where $\psi_N$ is normalized poloidal flux. It is calculated using SURFMN[43] to spectrally decompose the MP components and M3D-C1[44] to compute the linear one-fluid plasma response. The applied RMPs were largely shielded and had a maximum amplitude of $\delta B_r/B_T = 4.4 \times 10^{-4}$, corresponding to an I-coil current of 5.4 kA. Stochastic regions were predicted outside $\rho = 0.96$,[14] and kinetic profiles showed no evidence of island formation. The odd parity I-coil configuration used to produce NRMPs also created a small resonant component with $\delta B_r/B_T \approx 5\times10^{-5}$.

The time evolution of an example shot from the experiment is shown in Fig. 2. During each $P_{LH}$ measurement period, the plasma was initially held in L-mode by applying maximum amplitude RMPs, which keeps $P_{LH}$ high, while using relatively low auxiliary heating: $P_{NBI} = 1.7$ MW [Fig. 2(a)] and $P_{ECH} = 1.2$ MW [Fig. 2(b)]. The RMP amplitude was then stepped down [Fig. 2(c)] at constant heating power until an L-H transition occurred at 1758 ms, as seen by the inflection point in $\bar{n}_e$ [Fig. 2(d)] and sharp decrease in recycling $D_\alpha$ light at the divertor outer strike point [Fig. 2(e)]. We chose this methodology, as opposed to keeping the RMP amplitude constant while stepping up heating power, because the RMP amplitude can be stepped with finer resolution. Following the RMP step-down, the auxiliary heating was turned off to induce an H-L back-transition (marked by a spike in $D_\alpha$),[45] allowing a second RMP step-down scan to be performed. The second scan had lower ECH power, 1.0 MW instead of 1.2 MW, so the second L-H transition at 3903 ms occurred at lower I-coil current than the first did (2.7 kA instead of 3.6 kA). RMPs also appear to alter the H-L back-transition power threshold, but the cause is unknown and is not a part of this investigation.

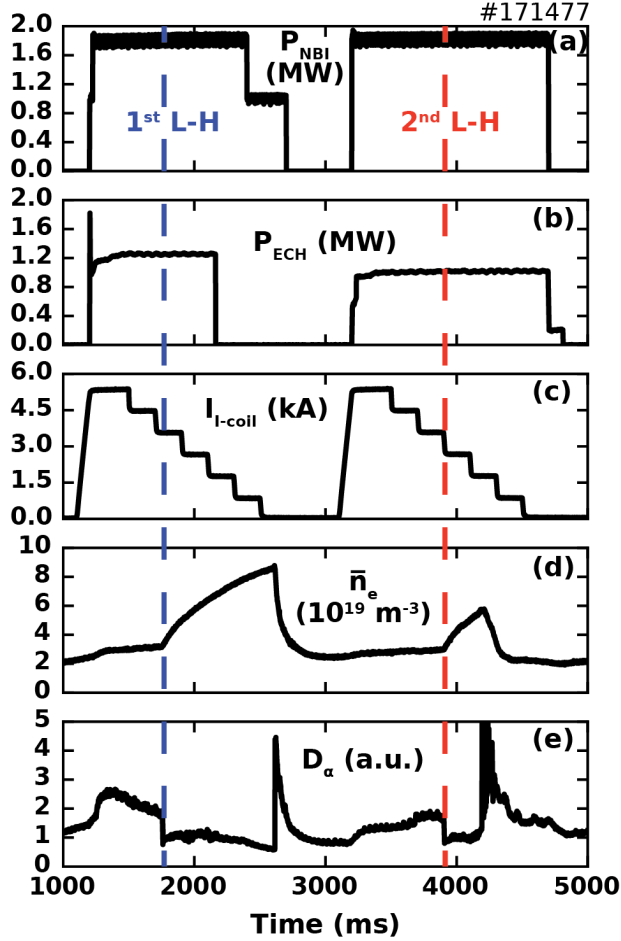

**FIG. 2.** Time evolution of (a) neutral beam injection power $P_{NBI}$, (b) electron cyclotron heating power $P_{ECH}$, (c) I-coil current $I_{I\text{-coil}}$, (d) line-averaged electron density $\bar{n}_e$, and (e) recycling D-alpha emission at the divertor outer strike point $D_\alpha$. The two L-H transitions, which occur at different $I_{I\text{-coil}}$ and $P_{ECH}$, are marked by dashed blue and red lines.

Density and flow fluctuation data was obtained using the beam emission spectroscopy (BES)[46,47] and Doppler backscattering (DBS)[48–50] diagnostics. BES measures spatially localized, long-wavelength ($k_\perp \rho_s \lesssim 0.5$) density fluctuations by viewing the Doppler-shifted $D_\alpha$ emission, i.e., the $n = 3$ to $n = 2$ atomic transition in deuterium, from heating neutral beam atoms.[51] In this experiment, the BES channels were configured in an 8×8 (radial by poloidal) array spanning $\rho =$ 0.85–1.04 [Fig. 1(a)], where $\rho$ is normalized toroidal flux. Each channel imaged a 1.0 cm (radial) × 1.5 cm (poloidal) area, but the effective radial width of each channel was broadened to 1.3 cm by excited state lifetime effects.[52] DBS measures spatially localized, intermediate-wavelength ($k_\theta \rho_s = 0.4$–$0.7$) density fluctuations by probing the plasma with a microwave beam and measuring the amplitude of the backscattered radiation. There were two DBS systems separated 180° toroidally with 8 radially separated channels each, spanning roughly the same radial range as BES.

In addition to density fluctuations, both BES and DBS measured the turbulence propagation velocity $v_{\text{turb}}$. For DBS the poloidal component of $v_{\text{turb}}$ was determined from the Doppler shift of the backscattered signal,[48] while for BES $v_{\text{turb}}$ was inferred using velocimetry analysis,[53] which utilizes the entire array of channels to track the motion of turbulent eddies as they propagate across the BES field of view. In this work, an image processing algorithm called orthogonal dynamic programming[54] was used to perform velocimetry analysis. Both the radial and poloidal components of $v_{\text{turb}}$ were inferred with 1 µs time resolution, fully resolving turbulent velocity fluctuations. Recent work has quantified the uncertainty in the inferred flow and shown that accurate results require signal-to-noise ratios $\gtrsim 10$, which was readily satisfied for the edge BES data from this experiment.[55]

The turbulence propagation velocity was measured in the lab frame and therefore satisfies $v_{\text{turb}} = v_{E \times B} + v_{\text{ph}}$,[56] where $v_{E \times B}$ is the $E \times B$ velocity and $v_{\text{ph}}$ is the turbulence phase velocity in the plasma frame, i.e., the reference frame where $E_r = 0$. For drift-wave modes $v_{\text{ph}}$ is a fraction of the diamagnetic velocity $v_D$, which was 3–8 km/s in the edge region of the plasmas in this experiment. It is frequently assumed that $v_{E \times B} \gg v_{\text{ph}}$, so $v_{\text{turb}} \simeq v_{E \times B}$, which is valid in the core of strongly rotating plasmas.[57] However, for this experiment $v_{\text{ph}}$ was comparable to $v_{E \times B}$ since measurements were made in the edge region of slowly rotating plasmas.

This has two implications. First, BES and DBS measure different $v_{\text{turb}}$ since $v_{\text{ph}}$ depends on $k$. For drift-wave modes $v_{\text{ph}}$ scales inversely with $k$,[58] so the DBS measurements of $v_{\text{turb}}$ are expected to be closer to $v_{E \times B}$ than for BES. Second, the shear in $v_{\text{turb}}$ is equal to the sum of the $E \times B$ shear and the $v_{\text{ph}}$ shear. The $E \times B$ shear has been extensively shown to suppress turbulence.[19] The effect of $v_{\text{ph}}$ shear on turbulence has been considered far less extensively than $E \times B$ shear, but it has been shown in some gyrokinetic simulations to play a comparable role in regulating transport.[59] Therefore, in this paper we use the full $v_{\text{turb}}$ shear instead of just $E \times B$ shear. Throughout the rest of this paper we drop the *turb* subscript, so $v$ refers to the turbulence propagation velocity.

We analyzed turbulence leading up to and across L-H transitions for three cases: no applied MPs (axisymmetric), maximum amplitude RMPs, and maximum amplitude NRMPs. Parameters for these cases are given in Table I. We selected the time windows by taking the longest period preceding each L-H transition with a stationary turbulence spectrum. All cases had $\bar{n}_e \approx 3 \times 10^{19}$ m$^{-3}$ and exhibited sharp L-H transitions. We note that at $\bar{n}_e \approx 1.5 \times 10^{19}$ m$^{-3}$ the L-H transitions instead exhibited a ~100 ms limit cycle oscillation phase, similar to that in Ref. 25. The power threshold is calculated using

$$P_{\text{LH}} = P_{\text{oh}} + P_{\text{NBI}} + P_{\text{ECH}} - \dot{W}, \tag{1}$$

where $P_{\text{oh}}$ is the ohmic heating power and $\dot{W}$ is the time rate of change of stored energy. The core radiated power is not subtracted because bolometry measurements for this experiment were dominated by scrape-off layer radiation. For the axisymmetric case, RMPs were initially applied before the analyzed time window to hold the plasma in L-mode and then turned off, so the calculated $P_{\text{LH}}$ value of 2.2 MW is an upper bound on the true value. Previous experiments using the same plasma conditions, but without RMPs, measured $P_{\text{LH}} = 2.1$ MW.[60] Given the ±0.1 MW variation in $P_{\text{LH}}$ from shot-to-shot due to slight changes in density and wall conditions, there is no significant difference in $P_{\text{LH}}$ for the axisymmetric and NRMP cases. In contrast, 0.7 MW of additional ECH power is required to access H-mode when RMPs are applied, so $P_{LH}$ is 30% higher. This increase cannot be attributed to RMPs enhancing fast ion losses, which is only predicted to cause about a 5% loss of NBI power.[61]

**TABLE I.** Parameters of analyzed L-H transitions for three cases with different applied MPs.

| MP type | None | Resonant | Non-resonant |
|---|---|---|---|
| Discharge # | 171472 | 171473 | 171495 |
| Time window (ms) | 2157–2257 | 2200–2525 | 3300–3468 |
| $\bar{n}_e$ ($10^{19}$ m$^{-3}$) | 3.2 | 3.1 | 3.0 |
| $P_{oh}$ (MW) | 1.0 | 0.8 | 1.1 |
| $P_{NBI}$ (MW) | 1.5 | 1.5 | 1.5 |
| $P_{ECH}$ (MW) | 0.3 | 1.0 | 0 |
| $\dot{W}$ (MW) | 0.6 | 0.5 | 0.5 |
| $P_{LH}$ (MW) | 2.2 | 2.8 | 2.1 |

## III. EFFECT OF MAGNETIC PERTURBATIONS ON L-MODE TURBULENCE AND FLOWS

This section presents how $n = 3$ MPs alter turbulence and flows in the stationary L-mode state preceding the L-H transition, using the time windows listed in Table I. We investigate how MPs affect edge temperature and density profiles, density fluctuations, flow profiles, shear rates, Reynolds stress, and turbulence correlation properties.

### A. Edge density and temperature profiles

Before investigating how MPs alter turbulence and flows, we examine how they affect profile gradients since these govern the linear instability drive for turbulence. Figure 3 shows L-mode electron density $n_e$ and temperature $T_e$ profiles measured by Thomson scattering and fit using OMFITprofiles.[62] Their normalized gradients, $\nabla n_e/n_e$ and $\nabla T_e/T_e$, are also calculated (defining $\nabla = \partial/\partial\rho$). Both RMPs and NRMPs reduce the peak value of $\nabla n_e/n_e$ by 30% [Fig.

3(c)]. RMPs reduce $\nabla T_e/T_e$ by 20%, primarily because the additional ECH power needed to induce an L-H transition raises $T_e$, but $\nabla T_e/T_e$ is not affected by NRMPs [Fig. 3(d)]. The edge profiles become toroidally modulated when MPs are applied,[63] but in these L-mode plasmas the modulation is small. The variation with MPs shown in Fig. 3 reflects the maximum change.[14]

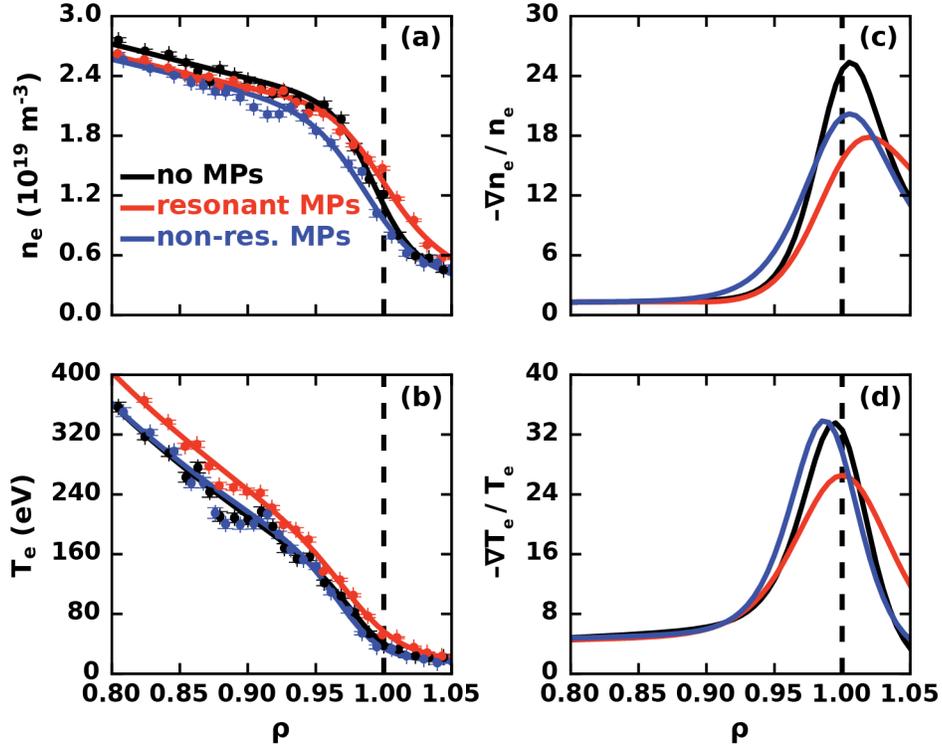

**FIG. 3.** Effect of MPs on edge profiles of (a) electron density $n_e$, (b) electron temperature $T_e$, (c) normalized $n_e$ gradient, and (d) normalized $T_e$ gradient. RMPs and NRMPs reduce the peak $n_e$ gradient by 30%. RMPs also reduce the peak $T_e$ gradient by 20%, while NRMPs have no effect.

The influence of these normalized gradients on linear instability drive is mode dependent. As will be discussed in Sec. III B, the edge turbulence is a mixture of modes that may include ion temperature gradient (ITG) modes, trapped electron modes (TEMs), and resistive ballooning modes (RBMs). For ITG modes $\nabla T_i/T_i$ is destabilizing, while $\nabla n_i/n_i$ is stabilizing.[58] Charge exchange recombination (CER) measurements of $T_i$ were not collected for these discharges, but previous experiments in similar plasmas found that RMPs had little effect on carbon $T_i$.[64] Carbon impurity concentrations were low ($Z_{eff}$ varied from 1.8 to 1.6 over $\rho = 0.90$ to 1.00), so $n_i \approx n_e$.

Assuming $\nabla T_i / T_i$ is unaffected by RMPs (as was found in Ref. 64), the reduction of $\nabla n_e / n_e$ with RMPs is therefore expected to enhance ITG drive. For TEMs, larger $\nabla n_e / n_e$ and $\nabla T_e / T_e$ are both destabilizing,[65] so RMPs have a stabilizing influence. For RBMs, growth rates scale with the total pressure gradient $\nabla p = \nabla p_i + \nabla p_e$,[66] so by reducing $\nabla n_e$ RMPs have a stabilizing influence, again assuming $\nabla T_i$ is largely unaffected.

**B. Density fluctuations**

Figure 4(a) shows density fluctuation power spectra measured by BES at $\rho = 0.93$. The uncertainty bands for this plot, and all others in this paper, reflect 95% confidence intervals (±1.96 standard deviations about the mean). Uncertainties are generally lowest for the RMP case since it has the longest time window. The spectra are Doppler shifted, i.e., the measured frequency is given by $\omega = \omega_\text{plasma} + k_\theta v_{E \times B}$, where $\omega_\text{plasma}$ is the turbulence frequency in the plasma frame and $k_\theta$ is the turbulence mean poloidal wavenumber.

Two turbulence modes are observed: a low-frequency mode peaking around 3 kHz and a high-frequency mode peaking around 35–55 kHz depending on the type of applied MP. These turbulence frequencies are somewhat lower than typical for DIII-D because balanced neutral beam injection was used to keep the rotation low, reducing the Doppler shift. Both modes propagate in the lab-frame electron diamagnetic direction [Fig. 4(b)] and have coherency significantly above the noise floor (dashed line) up to 115–150 kHz depending on MP type [Fig. 4(c)]. RMPs raise the normalized root-mean-square density fluctuation amplitudes $\tilde{n}_\text{rms} / \bar{n}$ of both modes by 20%, while NRMPs damp them by 10%. RMPs also reduce the Doppler shift of the high-frequency mode by damping the plasma's toroidal rotation.

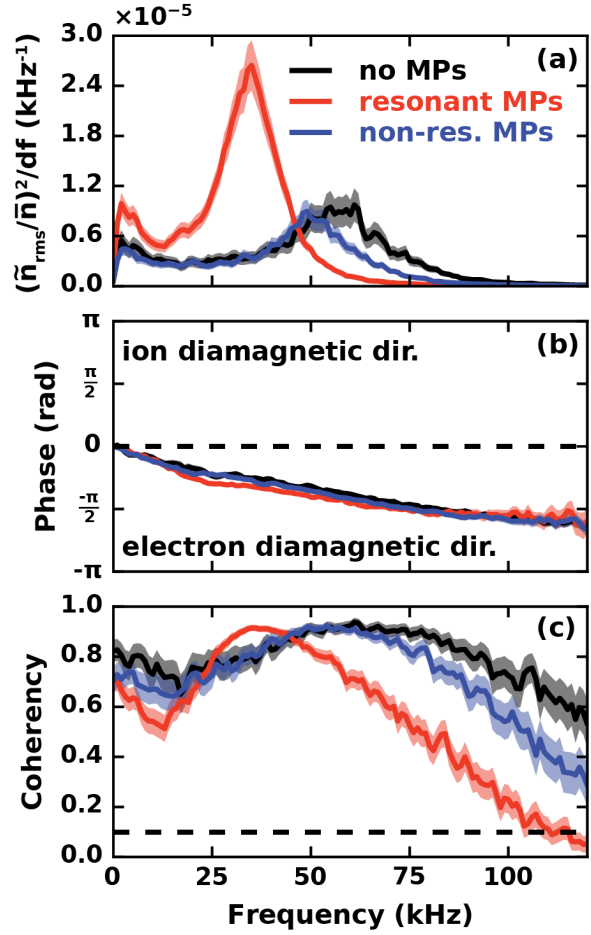

**FIG. 4.** Impact of MPs on density fluctuation (a) power spectra, (b) cross-phase spectra, and (c) coherency spectra measured by BES at $\rho = 0.93$ using poloidally adjacent channels. The dashed line in (c) denotes the coherency noise floor. Two modes propagating in the lab-frame electron diamagnetic direction are present, peaking respectively around 3 kHz and 35–55 kHz. Both are amplified by RMPs but slightly damped by NRMPs.

Density fluctuation spectra are largely unaffected when the toroidal phase of the RMP is rotated by 60º, as shown by Fig. 5. These measurements are from different plasmas than the rest of this paper that have identical parameters but only 80% of the full RMP amplitude applied ($\delta B_r/B_T = 3.5\times 10^{-4}$), which is still large enough to raise $P_{LH}$. The BES channels are toroidally located near the center of an I-coil (where the RMP amplitude peaks), so the fact that there is no significant change in the turbulence spectra when the RMP is rotated indicates the turbulence is largely axisymmetric in the edge of these L-mode plasmas. This is in contrast to turbulence in H-mode pedestals with applied MPs, which has significant 3D variation.[63,67]

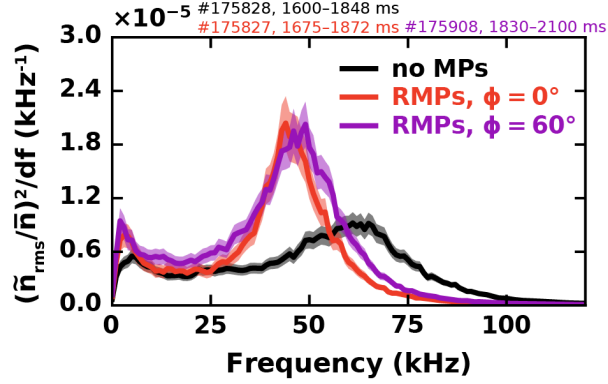

**FIG. 5.** Impact of RMP toroidal phase on density fluctuation power spectra measured by BES at $\rho = 0.93$. The spectra do not significantly change when the RMP toroidal phase is rotated by 60°, indicating the turbulence is largely axisymmetric.

The radial dependence of $\tilde{n}_{rms}/\bar{n}$ for each mode is shown in Fig. 6. The fluctuation amplitude $\tilde{n}_{rms}$ is calculated by taking the square root of the integrated power spectra over frequency ranges where the coherency is above the noise floor. The frequency separating the two modes is determined using the sharp inflection point in coherency, e.g., 10–15 kHz in Fig. 4(c). Due to the long time-averaging windows employed and high signal-to-noise ratio for L-mode edge data, the random uncertainty in $\tilde{n}_{rms}/\bar{n}$ is at the 0.1% level and errorbars are not visible in Fig. 6. Note that the effect of MPs on $\tilde{n}_{rms}/\bar{n}$ appears less pronounced in Fig. 6 than in Fig. 4(a) because the power spectra in Fig. 4(a) are proportional to $(\tilde{n}_{rms}/\bar{n})^2$.

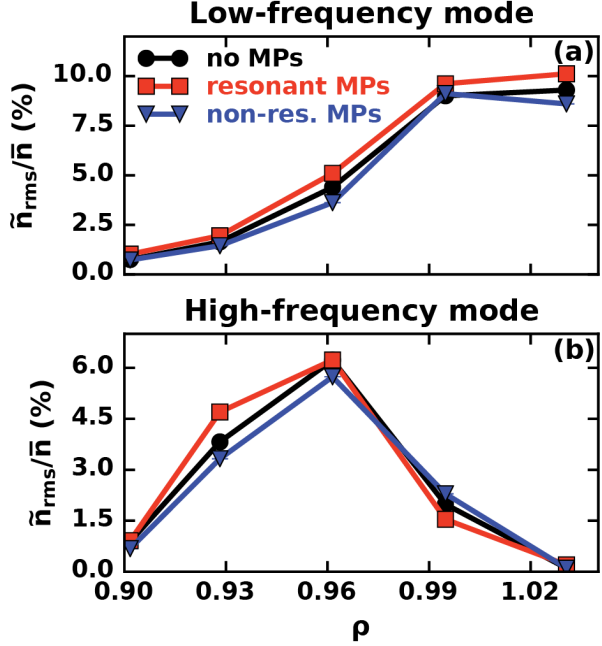

**FIG. 6.** Effect of MPs on density fluctuation amplitudes for (a) the low-frequency mode, likely an ion temperature gradient mode, and (b) the high-frequency mode, likely a trapped electron mode. The low-frequency mode peaks in amplitude at the separatrix and is slightly amplified by RMPs over $\rho = 0.93$–$1.04$. The high-frequency mode peaks at $\rho = 0.96$ and is amplified by RMPs only at $\rho = 0.93$.

The low-frequency mode peaks in amplitude at the separatrix and is amplified by RMPs by 20% on average over $\rho = 0.93$–$1.04$ [Fig. 6(a)]. The high-frequency mode peaks in a narrow range around $\rho = 0.96$ [Fig. 6(b)]. RMPs amplify it by 20% at $\rho = 0.93$, damp it by 20% at $\rho = 1.00$, and have no effect elsewhere. In general, NRMPs have the opposite effect on $\tilde{n}_{\rm rms}/\bar{n}$ as RMPs.

The direction of the plasma-frame turbulence phase velocity $v_{\rm ph}$ can be used to help identify each mode. CER data to determine $v_{E\times B}$ could not be collected at the same time as BES data, so the lab-frame turbulence velocity $v$ measured by BES cannot be directly transformed to the plasma frame. However, at the separatrix $v_{E\times B}$ is expected to be near zero, so $v \approx v_{\rm ph}$. There, the low-frequency mode propagates in the ion diamagnetic direction and the high-frequency modes propagate in the electron diamagnetic direction. The low-frequency mode also has lower $k_\theta$ than the high-frequency mode, since the spectra are Doppler shifted. These observations suggest that the low-frequency mode is an ITG mode and the high-frequency mode is a TEM or RBM.

Previous simulation work on DIII-D L-mode plasmas with similar shapes and edge parameters supports the notion that the edge turbulence is a mix of ITG modes, TEMs, and/or RBMs. Linear stability calculations using TGLF[68] found that either ITG modes or TEMs were unstable, depending on RMP amplitude.[64] Nonlinear gyrokinetic simulations using GENE[69] showed a hybrid state of ITG modes and TEMs.[70] Nonlinear fluid simulations using BOUT++[71] found that RBMs were unstable and produced a dual, counter-propagating mode structure similar to that in experiment, while ITG modes and TEMs were stable.[72] Gyrokinetic codes and BOUT++ generally disagree about which modes are unstable in the edge of DIII-D L-mode plasmas, and more work is needed to resolve this disagreement.

The observed changes in $\tilde{n}_{\rm rms}/\bar{n}$ with applied MPs (Fig. 6) are not consistent with the changes in linear instability drive for ITG modes, TEMs, and RBMs expected from the reduction of $\nabla n_e/n_e$ and $\nabla T_e/T_e$ (Fig. 3), assuming $\nabla T_i/T_i$ is unaffected. ITG drive is expected to go up with both RMPs and NRMPs, but $\tilde{n}_{\rm rms}/\bar{n}$ instead goes up with RMPs and down with NRMPs throughout the edge region. Linear drive for both TEMs and RBMs is expected to go down with both RMPs and NRMPs, but $\tilde{n}_{\rm rms}/\bar{n}$ instead exhibits a non-monotonic radial dependence and changes in opposite directions for RMPs and NRMPs. Even with the uncertainty about which modes the turbulence consists of, these results imply that changes in profile gradients do not explain the observed turbulence modifications with applied MPs.

**C. Turbulence flow profiles and shear rates**

To investigate how MPs affect flows, we apply velocimetry analysis to the BES data. The density fluctuation data is first bandpass filtered to remove photon, electronic, and beam-power-supply noise. The bandpass cutoff frequencies vary with radius and type of MP. They are set to values where the coherency between poloidally adjacent channels is above the noise floor. Both the low- and high-frequency modes are passed, so the inferred $v$ is an amplitude-weighted average of the two modes' velocities. To obtain numerically converged velocimetry results, the filtered, 8×8 density fluctuation images are spatially interpolated to 40×40 resolution using cubic radial basis functions. The inferred velocity fields are then downsampled back to the original 8×8 resolution. Mean flows are calculated by averaging the flow-field in time and the poloidal direction. Random uncertainties are estimated by computing standard errors of the time-averaged flow over the poloidal direction.

Figure 7 shows how MPs alter the mean turbulence poloidal velocity $\bar{v}_\theta$, using measurements from BES and DBS. Both BES [Fig. 7(a)] and DBS [Fig. 7(c)] show a velocity well in the electron diamagnetic direction centered around $\rho = 0.95$ with a depth of 2–7 km/s. This is about an order of magnitude smaller than the depth of the well in a fully developed H-mode. The BES data shows deeper $\bar{v}_\theta$ wells than DBS [note the different y-axis scales for Fig. 7(a) and 6(c)], consistent with the higher sensitivity of BES to low-$k$ turbulence with larger phase velocity. The $\bar{v}_\theta$ well is also wider with BES due to the larger width of the BES channels, which is represented by the horizontal bar in Fig. 7(a) at $\rho = 0.96$. RMPs reduce the depth of the $\bar{v}_\theta$ well by 60% compared to the axisymmetric case. This percentage reduction is the same for BES and DBS, even though the absolute $\bar{v}_\theta$ well depths measured by each diagnostic differ. NRMPs also reduce the $\bar{v}_\theta$ well depth, but by only 20–30%.

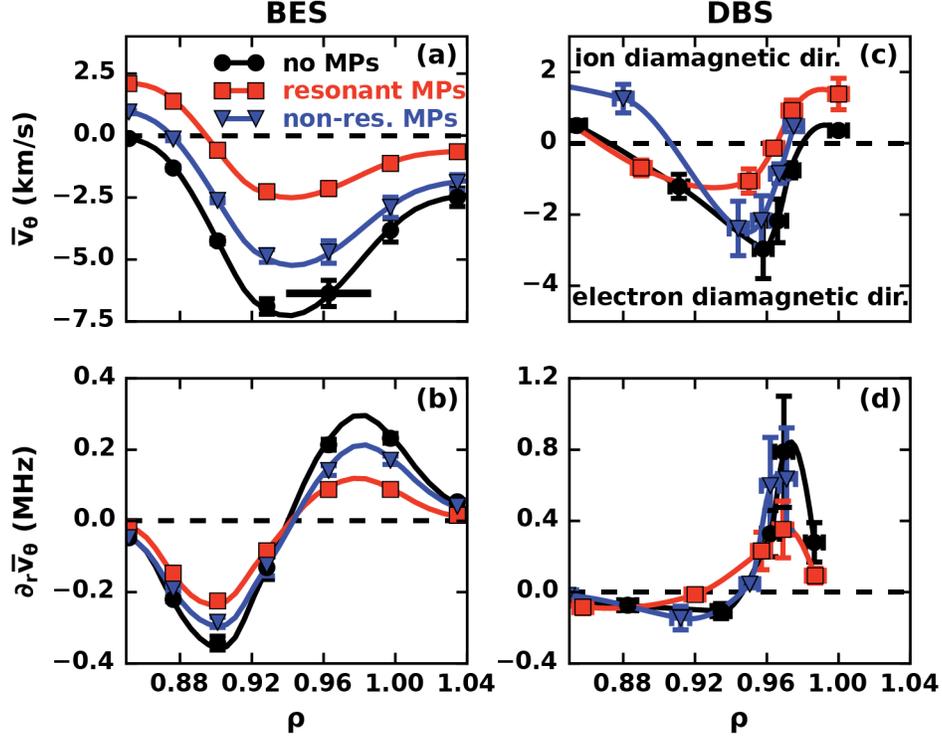

**FIG. 7.** Effect of MPs on profiles of (a,c) mean turbulence poloidal velocity $\bar{v}_\theta$ and (b,d) its shear rate $\partial_r \bar{v}_\theta$ measured by (a,b) BES and (c,d) DBS. The horizontal bar at $\rho = 0.96$ in (a) represents the width of a BES channel. Both diagnostics show that RMPs, and to a lesser extent NRMPs, reduce the $\bar{v}_\theta$ well depth and correspondingly reduce the shear rate in the outer shear layer.

Flow shear rates given by $\partial_r \bar{v}_\theta$ are shown in Fig. 7(b,d). The two sides of the $\bar{v}_\theta$ well form the inner shear layer, near $\rho = 0.90$, and outer shear layer, near $\rho = 0.97$. Both BES and DBS show that RMPs reduce the shear rate at the outer shear layer by 60% while NRMPs reduce it by only 20%. DBS shows larger shear rates at the outer shear layer than the inner shear layer, as opposed to BES, which shows a more symmetric $\bar{v}_\theta$ well structure with near equal shear rates at the inner and outer shear layers. Turbulence suppression during the L-H transition is typically observed to start at the outer shear layer and then propagate inward, so we focus on the shear modifications there. The peak DBS shear rates at the outer shear layer are about three times larger than those measured by BES. However, the percentage reductions in shear rates with applied MPs are the same for both diagnostics.

Previous experiments have also observed that RMPs reduce the $\bar{v}_\theta$ well depth and shear rate when applied in L-Mode.[10,14] In particular, recent analysis of the same DIII-D plasmas used here has attributed the reduction in shear rate to magnetic stochasticity created by the RMP.[14] In this model, magnetic islands formed by the RMP overlap and stochasticize the magnetic field for $\rho > 0.97$. The fast, parallel motion of electrons along the stochastic field lines then produces a radial electron current $j_r^e$ that shorts out $E_r$, reducing $v_{E \times B}$ and hence the depth of the $\bar{v}_\theta$ well. This model is consistent with RMP results but does not as readily explain how NRMPs, which produce substantially less island overlap and stochasticity, reduce the $\bar{v}_\theta$ well depth. In Sec. III E we will show that both resonant and non-resonant MPs reduce Reynolds stress flow drive, contributing to the reduction of $\bar{v}_\theta$ and potentially explaining this discrepancy.

### D. Velocity fluctuations

Figure 8 shows $v_\theta$ fluctuation power spectra at $\rho = 0.88$ and $\rho = 0.96$. RMPs excite a geodesic acoustic mode (GAM) over $\rho = 0.85-0.90$, most strongly at $\rho = 0.88$ [Fig. 8(a)]. The GAM is the large peak in the spectrum at 17 kHz. This peak is thought to be a GAM because the velocity fluctuations are in phase over the poloidal extent of the BES array, consistent with the $m = 0$ property of GAMs, and the frequency matches the range of GAM frequencies previously observed in detailed investigations on DIII-D.[73] A smaller amplitude GAM is also observed when NRMPs are applied, but there is no evidence of GAMs in the axisymmetric case.

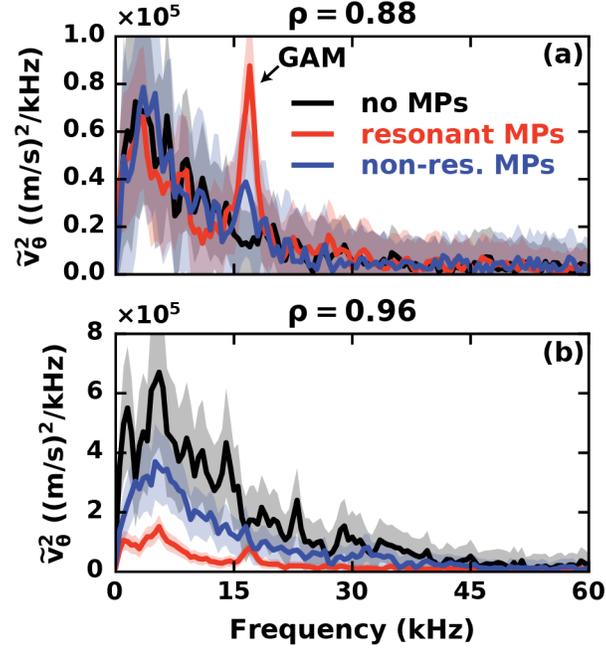

**FIG. 8.** Effect of MPs on power spectra of turbulence poloidal velocity fluctuations $\tilde{v}_\theta$ at (a) $\rho = 0.88$ and (b) $\rho = 0.96$. RMPs excite a geodesic acoustic mode (GAM) at $\rho = 0.88$ and reduce broadband $\tilde{v}_\theta$ amplitudes at $\rho = 0.96$.

RMPs exciting a GAM is contrary to both theoretical expectations and previous measurements on other machines. Several theories predict RMPs to increase the damping rate of zonal flows.[34,35] These theories strictly apply to low-frequency zonal flows, but are thought to also apply to GAMs as the two are closely related. Consistent with these theories, measurements on TEXTOR and MAST have shown that GAM amplitudes decrease when RMPs are applied.[74,75] The measurements here on DIII-D imply that RMPs do not always damp GAMs, challenging these theories. Turbulence suppression across the L-H transition primarily occurs at $\rho = 0.95-1.00$ and no GAMs are observed for $\rho \geq 0.90$, so GAMs do not appear to play a role in triggering the L-H transition. Nevertheless, these results are of theoretical interest due to the connection between GAMs and low-frequency zonal flows, which do play an important role in L-H transition dynamics.

In the edge region, $\rho = 0.90–1.00$, RMPs reduce $\tilde{v}_\theta$ amplitudes, as illustrated by the spectra at $\rho = 0.96$ in Fig. 8(b). The spectra broadly peak around 5 kHz and monotonically decay to the noise floor at 100 kHz. The root-mean-square $\tilde{v}_\theta$ amplitude is reduced 55% by RMPs and 25% by NRMPs. This result is somewhat surprising since we previously showed that RMPs *raise* $\tilde{n}_{rms}/\bar{n}$. RMPs therefore have opposite effects on the density and velocity fluctuation fields. The broad low-frequency peak in the flow spectra is consistent with a low-frequency zonal flow. Its frequency is on the same order as the ion-ion collision frequency, consistent with the expectation that the zonal flow frequency is set by ion-ion collisions.[76] However, non-zonal velocity fluctuations are present over the same frequency band, making it difficult to separate the zonal and non-zonal components of the flow.

**E. Reynolds stress**

Reynolds stress is the fluctuation-driven momentum flux and is the mechanism by which turbulent flow fluctuations can affect mean flows. The full Reynolds stress tensor is defined as $\langle \tilde{\mathbf{v}}\tilde{\mathbf{v}} \rangle$, where $\tilde{\mathbf{v}}$ is the fluctuating component of the velocity and $\langle \ \rangle$ denotes a flux surface average. The divergence of the Reynolds stress tensor is a force on the mean flow and is called the Reynolds force. In the edge of tokamak plasmas, the Reynolds stress is dominated by the $\langle \tilde{v}_r \tilde{v}_\theta \rangle$ term and the Reynolds force then becomes $-\partial_r \langle \tilde{v}_r \tilde{v}_\theta \rangle$.[77] Both $\tilde{v}_r$ and $\tilde{v}_\theta$, as well as their radial derivatives, are inferred by velocimetry analysis of BES data. Since measurements are only made at the outboard midplane, we calculate the Reynolds stress using an average over the poloidal extent of the BES array instead of a flux surface average. We estimate the random uncertainty using the standard error of the poloidal average.

Figure 9 shows how MPs affect Reynolds stress and Reynolds force throughout the L-mode edge region. Reynolds stress peaks positively in the outer shear layer (near $\rho = 0.96$) and negatively in the inner shear layer (near $\rho = 0.90$) [Fig. 9(a)]. This produces a large, negatively directed Reynolds force at $\rho = 0.93$ [Fig. 9(b)], near the bottom of the $\bar{v}_\theta$ well [Fig. 7(a)]. The negative sign of the Reynolds force means the force is in the electron diamagnetic direction, consistent with the direction of the $\bar{v}_\theta$ well. This suggests the Reynolds force may make a significant contribution to the mean flow.

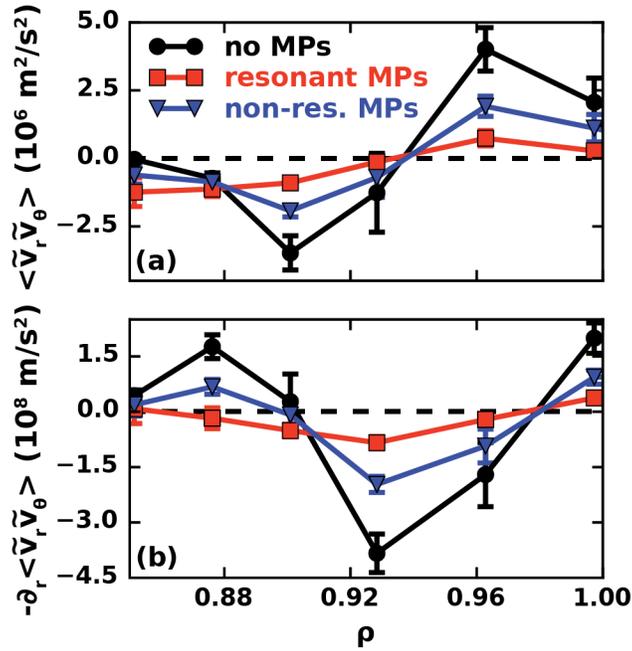

**FIG. 9.** Impact of MPs on (a) Reynolds stress $\langle \tilde{v}_r \tilde{v}_\theta \rangle$ and (b) Reynolds force $-\partial_r \langle \tilde{v}_r \tilde{v}_\theta \rangle$. RMPs reduce the amplitude of the Reynolds stress throughout the edge region, degrading the Reynolds force, which drives $\bar{v}_\theta$.

RMPs substantially reduce Reynolds stress over most of the edge region, degrading the Reynolds force at $\rho = 0.93$ by 80%. NRMPs degrade the Reynolds force by 55%. The Reynolds stress decreases with RMPs primarily due to the reduction of $\tilde{v}_\theta$ shown in Sec. III D, although there are additional 20% reductions in $\tilde{v}_r$ and the cross-phase between $\tilde{v}_r$ and $\tilde{v}_\theta$. The substantial reduction in Reynolds force at $\rho = 0.93$ is consistent with the reduction in $\bar{v}_\theta$ well depth shown in Sec. III C and may even cause it. The more modest reduction of Reynolds force with NRMPs is also consistent with the modest reduction in $\bar{v}_\theta$ well depth in that case.

To quantify the effect of reduced Reynolds force on $\bar{v}_\theta$, we employ a simplified version of the poloidal momentum balance equation that neglects all terms except Reynolds force and neoclassical flow damping. The equation, obtained by combining Eq. 22 and Eq. 28 in Ref. 77 and setting the time derivative term to zero, is

$$-\partial_r \langle \tilde{v}_r \tilde{v}_\theta \rangle = \gamma_{\text{damp}} \left( \frac{B_\phi}{B_\theta} \right)^2 \bar{v}_\theta, \tag{2}$$

where $\gamma_{\text{damp}}$ is the neoclassical poloidal flow damping rate. At the bottom of the $\bar{v}_\theta$ well, RMPs reduce Reynolds force by $3.0\times10^8$ m/s$^2$ and $\bar{v}_\theta$ by 4.6 km/s. At the outboard midplane $B_\phi/B_\theta = 3.5$ and $\gamma_{\text{damp}} = 0.9\times10^3$ s$^{-1}$, where $\gamma_{\text{damp}}$ is evaluated using Eq. B19 in Ref. 78, which accounts for multi-collisionality effects in the near-separatrix magnetic geometry of these plasmas. Plugging these values for $B_\phi/B_\theta$ and $\gamma_{\text{damp}}$ into Eq. 2, the predicted amount of Reynolds force reduction needed to produce the observed amount of $\bar{v}_\theta$ reduction is $0.5\times10^8$ m/s$^2$. The observed reduction in Reynolds force is six times larger than predicted, implying that Reynolds force is quantitatively large enough to significantly modify poloidal flow.

One potential cause for this discrepancy is that the terms in the theory behind Eq. 2 are flux surface averages while the experimental data is averaged over a narrow region at the outboard midplane. Fluctuation amplitudes are largest at the outboard midplane, so Reynolds force is also expected to be largest there. The flux-surface-averaged Reynolds force is therefore smaller than the outboard midplane value, although the exact difference is not known because Reynolds stress measurements off the outboard midplane are not possible on DIII-D. In addition, there are neglected terms in the poloidal momentum balance equation that likely change with RMPs but cannot currently be measured, e.g., Maxwell stress. Despite these caveats, this simplified poloidal flow model supports the conclusion that RMP-induced Reynolds force degradation plays a role in reducing the $\bar{v}_\theta$ well depth.

As discussed in Sec. III C, previous research has attributed the reduction in $\bar{v}_\theta$ well depth to stochasticity-induced $j_r^e$.[14] Stochasticity also offers a potential explanation for the reduction in Reynolds force shown here. Turbulence must be radially asymmetric to produce finite Reynolds force,[77] and $E\times B$ shear provides such an asymmetry. As stochasticity-induced $j_r^e$ shorts out $E_r$, the resulting reduction in $E\times B$ shear makes the $\bar{v}_\theta$ well shallower. This reduces the turbulence's radial asymmetry, causing the Reynolds force to go down and further shallowing the $\bar{v}_\theta$ well. In this way stochasticity-induced $j_r^e$ and reduced Reynolds force may act synergistically to reduce the $\bar{v}_\theta$ well depth. While $j_r^e$ scales inversely with collisionality, consistent with the $P_{\text{LH}} \sim (\nu^*)^{-0.3}$ scaling when RMPs are applied,[14] the collisionality scaling of Reynolds force is unknown.

**F. Turbulence correlation properties**

We now investigate how MPs affect more fundamental turbulence properties, including the correlation time $\tau_c$ and correlation lengths in the radial and poloidal directions $L_r$ and $L_\theta$. The correlation time is a measure of how long on average a turbulent eddy exists before differential advection of fluid parcels within the eddy causes it to decorrelate. Similarly, the correlation length is a characteristic size scale of the eddies, which in general is different in the radial and poloidal directions.

Correlation analysis between multiple pairs of BES channels is used to measure $\tau_c$, $L_r$, and $L_\theta$, as described in detail in Ref. 79. For each column in the 8×8 BES array, one channel, denoted the reference channel, is cross-correlated with itself and the other seven channels within its column.

The correlation functions are normalized so that 1 corresponds to perfect correlation and -1 to perfect anticorrelation. Figure 10(a) shows three of the eight correlation functions at $\rho = 0.96$ for the high-frequency mode in the axisymmetric case. The correlation functions (solid lines) oscillate, indicating the turbulence has wave-like behavior, and both shift in time and decrease in amplitude as the channel separation increases. The envelope of each correlation function (dashed lines) is calculated using the Hilbert transform to give a measure of the total correlation (both positive and negative correlation).

To extract $\tau_c$ the maximum value of each correlation envelope is plotted against the time-lag where the maximum occurs [red arrows in Figs. 9(a) and 9(b)]. The correlation typically follows a Gaussian decay, and $\tau_c$ is determined from the $1/e$ decay time of a Gaussian fit [the solid line in Fig. 10(b); the dashed line denotes the $1/e$ level]. To extract $L_\theta$ the value of the correlation envelope at zero time-lag is plotted against poloidal separation distance [purple arrows in Figs. 9(a) and 9(c)], and another Gaussian fit is performed [Fig. 10(c)]. $L_r$ is measured in the same way as $L_\theta$ except radially separated channels are used.

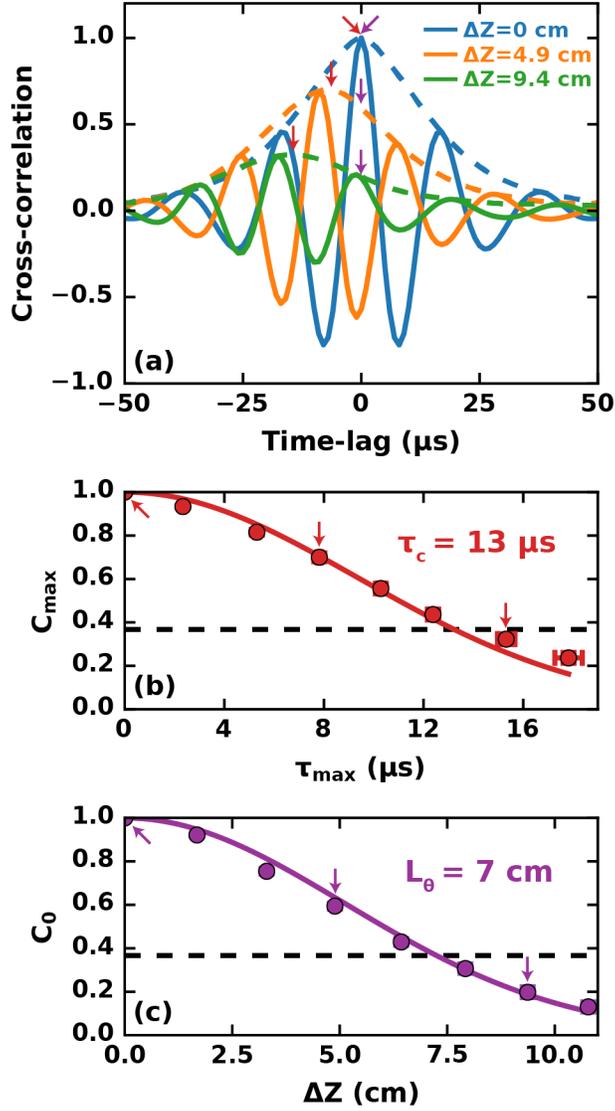

**FIG. 10.** (a) Cross-correlation functions (solid lines) and their envelopes (dashed lines) between poloidal separated BES channels at $\rho = 0.96$ for the axisymmetric case. (b) Maximum value of the cross-correlation envelope vs time-lag where the maximum occurs, used to calculate the turbulence correlation time $\tau_c$. (c) Value of the cross-correlation envelope at zero time-lag vs poloidal separation between channels, used to calculate the turbulence poloidal correlation length $L_\theta$.

This procedure is repeated using different reference channels to obtain several statistically independent estimates of the correlation parameters at each radial location. These estimates are averaged together, and their random uncertainty is estimated by computing the standard error of the mean. This entire process is then repeated for each radial location. For the data from this experiment, signal-to-noise ratios are only high enough to measure correlation parameters in region of $\rho = 0.90$–$1.00$. In addition, only the correlation parameters of the high-frequency mode could be measured in detail. The low-frequency mode has large $L_\theta$, of similar size as the BES array, causing the time-shifts between all the correlation functions to be near zero and preventing an accurate correlation decay from being measured.

Profiles of $L_\theta$ and the turbulence decorrelation rate $\Delta\omega_D = 1/\tau_c$ for different types of applied MPs are shown in Fig. 11. MPs have a weak effect on $L_\theta$, with RMPs and NRMPs raising it by 20% and 10%, respectively, near the inner shear layer ($\rho = 0.90$–$0.93$) and having no effect elsewhere [Fig. 11(a)]. MPs also do not affect $L_r$, which is ≈2.8 cm throughout the edge region. RMPs substantially raise $\Delta\omega_D$ by 60% on average throughout the edge region [Fig. 11(b)]. In contrast, NRMPs have no statistically significant effect on $\Delta\omega_D$.

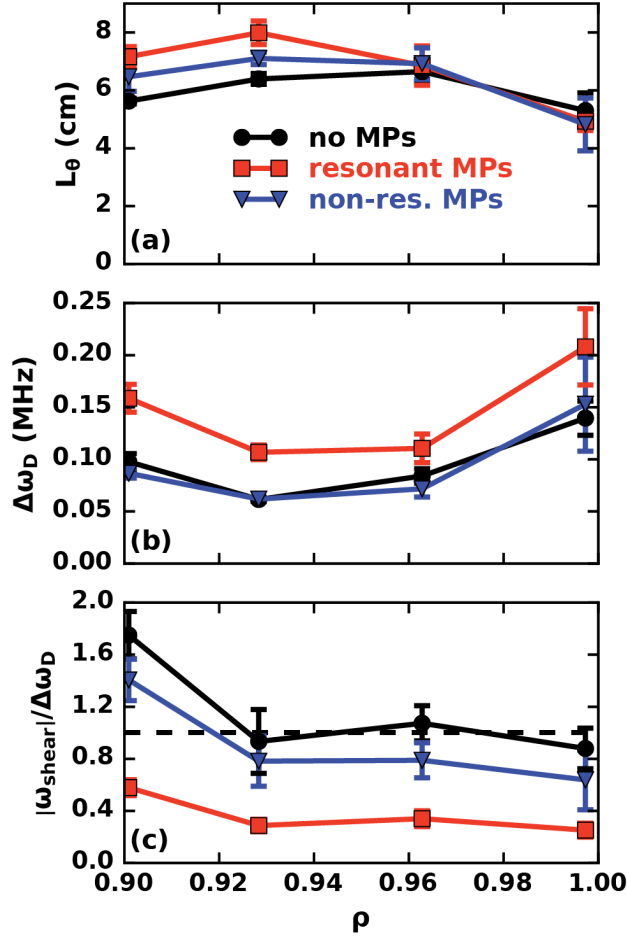

**FIG. 11.** Impact of MPs on (a) turbulence poloidal correlation length $L_\theta$, (b) turbulence decorrelation rate $\Delta\omega_D$, and (c) turbulence suppression parameter $\omega_{shear}/\Delta\omega_D$ for the high-frequency mode. RMPs raise $\Delta\omega_D$, reducing $\omega_{shear}/\Delta\omega_D$ to below 1 (marked by the dashed line) and therefore disrupting the turbulence shear suppression mechanism.

RMPs therefore modify turbulence so as to enhance the rate at which eddies decorrelate. This effect is independent of any shear flow modification of the turbulence: the observed decrease in shear rate would be expected to *reduce* $\Delta\omega_D$, not increase it as observed. The mechanism by which RMPs raise $\Delta\omega_D$ is currently unknown. One hypothesis, inspired by the non-zonal transition mechanism in gyrokinetic simulations,[80] is that RMPs cause magnetic field lines, and the turbulent eddies aligned to them, to radially drift from the unperturbed flux surfaces. The eddies can then irreversibly interact with neighboring regions of strong turbulence, causing them to decorrelate from the turbulence on the original flux surface and thereby raise $\Delta\omega_D$. Future investigations using nonlinear gyrokinetic simulations with RMPs applied could test this hypothesis and may otherwise help elucidate the mechanism by which $\Delta\omega_D$ increases.

To evaluate how turbulence suppression by shear flow is affected by MPs we must evaluate the effective shear rate $\omega_{shear}$ and compare it to $\Delta\omega_D$. The $E \times B$ shear rate in toroidal geometry is given by the Hahm-Burrell formula (Eq. 2 in Ref. 19):

$$\omega_{E \times B} = \frac{\Delta\psi}{\Delta\phi} \frac{\partial}{\partial \psi}\left(\frac{E_r}{RB_\theta}\right), \qquad (3)$$

where $\Delta\psi = L_r R B_\theta$ is the radial correlation length in poloidal flux units, $R\Delta\phi$ is the toroidal correlation length, $\psi$ is the poloidal flux, $E_r$ is the radial electric field, $R$ is the major radius, and $B_\theta$ is the poloidal magnetic field. For flute-like modes $L_\theta = RB_\theta \Delta\phi / B_\phi$. Substituting this into Eq. 3 and neglecting flux expansion terms, which are small compared to the radial variation of $v_{E \times B}$ at the outboard midplane, yields

$$\omega_{E \times B} = \frac{L_r}{L_\theta} \frac{\partial}{\partial r}\left(\frac{E_r}{B_\phi}\right). \qquad (4)$$

Replacing $v_{E \times B, \theta} = E_r/B_\phi$ with the turbulence velocity $\bar{v}_\theta$ gives an expression for the effective shear rate that can be calculated with experimentally measured quantities:

$$\omega_{shear} = \frac{L_r}{L_\theta} \partial_r \bar{v}_\theta. \qquad (5)$$

As explained in Sec. II, this shear rate expression includes both $E \times B$ shear and $v_{\text{ph}}$ shear.

Turbulence suppression by shear flow is parametrized by the ratio $\omega_{shear}/\Delta\omega_D$. Note that $\Delta\omega_D$ is the measured decorrelation rate in the presence of shear flow, so as $\omega_{shear}$ increases, $\omega_{shear}/\Delta\omega_D$ asymptotes to ~1. Figure 11(c) shows how MPs impact the shear suppression parameter. In the axisymmetric case, $\omega_{shear}/\Delta\omega_D \gtrsim 1$ throughout the entire edge region, implying the shear flow is suppressing the turbulence amplitude below its ambient level. When RMPs are applied, the combination of reduced $\omega_{shear}$ and increased $\Delta\omega_D$ causes $\omega_{shear}/\Delta\omega_D$ to decrease to significantly below 1, implying the shear flow no longer has a suppressing effect on the turbulence. The reduction of $\omega_{shear}$ is caused primarily by the reduction of $\partial_r \bar{v}_\theta$ as there is little change in $L_r$ or $L_\theta$ with MPs. NRMPs slightly reduce $\omega_{shear}$ and do not affect $\Delta\omega_D$, so $\omega_{shear}/\Delta\omega_D$ only decreases to slightly below 1.

These results show that RMPs disrupt the turbulence shear suppression mechanism in the stationary L-mode state preceding the L-H transition. In the edge of an H-mode plasma, the shear rate is large, so $\omega_{shear}/\Delta\omega_D \sim 1$ and the turbulence is suppressed. In the axisymmetric L-mode case investigated here, $\omega_{shear}/\Delta\omega_D$ is already ~1, so any mechanism that drives additional shear flow, even transiently, will further suppress the turbulence and trigger the L-H transition. When RMPs are applied $\omega_{shear}/\Delta\omega_D < 1$, so substantially more transient turbulence suppression is needed to trigger the L-H transition. This makes transitioning to H-mode more difficult and is consistent with the $P_{LH}$ increase when RMPs are applied.

NRMPs only slightly disrupt the turbulence suppression mechanism, so we expect them to slightly raise $P_{LH}$. In this experiment, there is no significant difference in $P_{LH}$ between the axisymmetric and NRMP cases. However, some previous experiments have indeed found that NRMPs slightly raise $P_{LH}$.[9]

## IV. EFFECT OF MAGNETIC PERTURBATIONS ON L-H TRANSITION TURBULENCE-FLOW DYNAMICS

In this section, we investigate how turbulence-flow dynamics on the ~100 μs timescale of the L-H transition are affected by MPs. We showed in Sec. III that RMPs disrupt turbulence suppression by shear flow in the stationary L-mode state preceding the L-H transition, so we expect more transient turbulence suppression at the L-H timescale will be needed when RMPs are applied. To investigate this, we evaluate how energy is exchanged between turbulence and flows during the L-H transition using BES measurements.

Figure 12 shows time traces of turbulence quantities at $\rho = 0.96$ over a 10 ms period leading up to and across L-H transitions with different applied MPs. This location is 1 cm inside the separatrix and is where turbulence suppression first occurs during the L-H transition. The transition time, denoted by vertical dashed lines, is determined using the sharp decrease in $v_\theta$ as a marker. All quantities are averaged using a 100 μs centered moving average, which is long enough to average over several turbulence correlation times ($\tau_c = 5$–$15$ μs) but short enough to resolve the fast dynamics across the transition.

Density fluctuation amplitudes rapidly drop on a ~100 μs timescale across the L-H transition [Fig. 12(a)]. They become so small after the transition that velocimetry analysis can no longer track the turbulent eddies, causing the inferred turbulence velocities to go to zero, while the true values are non-zero. The radial turbulence velocity $v_r$ is negative (radially inward) on average leading up to the transition [Fig. 12(b)]. The poloidal turbulence velocity $v_\theta$ is also negative (electron diamagnetic direction) on average and increases in magnitude as the transition is approached [Fig. 12(c)]. The Reynolds stress $\langle \tilde{v}_r \tilde{v}_\theta \rangle$ exhibits large, positively skewed bursts leading up to the transition [Fig. 12(d)]. In all three cases, a large Reynolds stress bursts occurs less than 600 μs prior to the transition. MPs reduce the amplitude of these bursts, with RMPs having a stronger effect than NRMPs. This suggests that MPs disrupt the Reynolds stress mechanism that is thought to play a key role in triggering the L-H transition.

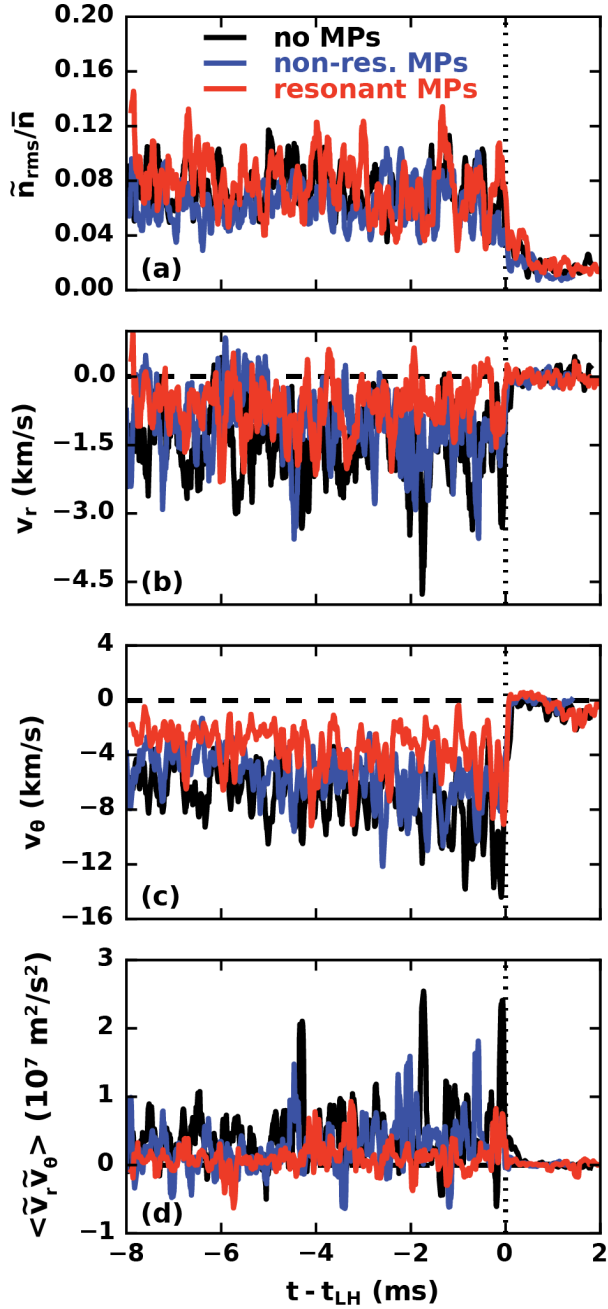

**FIG. 12.** Time traces leading up to the L-H transition at $\rho = 0.96$ (outer shear layer) of (a) normalized density fluctuation amplitude $\tilde{n}_{rms}/\bar{n}$, (b) radial turbulence velocity $v_r$, (c) poloidal turbulence velocity $v_\theta$, and (d) Reynolds stress $\langle \tilde{v}_r \tilde{v}_\theta \rangle$. Bursts of Reynolds stress are observed leading up to the L-H transition. RMPs substantially reduce the amplitude of these bursts while NRMPs reduce them less so.

To investigate this Reynolds stress reduction in more detail we evaluate the energy balance between turbulence and mean flows. We utilize a version of the predator-prey model described in Ref. 81 that nonlinearly conserves a total free energy that is composed of thermal free energy, turbulent flow energy, and mean flow energy. The thermal free energy density is given by

$$E_n = \frac{1}{2} n_0 T_{e0} \left(\frac{\tilde{n}}{n_0}\right)^2, \qquad (6)$$

the turbulent flow energy density by

$$E_\sim = \frac{1}{2} n_0 m_i \left(\langle \tilde{v}_r^2 \rangle + \langle \tilde{v}_\theta^2 \rangle\right), \qquad (7)$$

and the mean flow energy density by

$$E_- = \frac{1}{2} n_0 m_i \langle v_\theta \rangle^2, \qquad (8)$$

where $n_0$ is the electron density, $T_{e0}$ is the electron temperature, $m_i$ is the ion mass, $\tilde{n}/n_0$ is the normalized density fluctuation amplitude, and $v_r/v_\theta$ is the radial/poloidal component of the $E \times B$ velocity. Each quantity is decomposed into a zonal component by taking a flux surface average (denoted by $\langle \ \rangle$) and a non-zonal component $\tilde{a} = a - \langle a \rangle$, for any quantity $a$.

In this model, energy is transferred nonlinearly from turbulent flows to mean flows at a rate given by

$$P = n_0 m_i \langle \tilde{v}_r \tilde{v}_\theta \rangle \partial_r \langle v_\theta \rangle, \qquad (9)$$

which is the Reynolds stress times the mean flow shear rate. The L-H transition is predicted to occur when $P$ becomes positive, indicating energy transfer *out of* turbulence and *into* flows, and large enough to deplete the total turbulence energy $E_n + E_\sim$. The *total* turbulence energy, not just $E_\sim$, must be depleted because energy is transferred between $E_n$ and $E_\sim$ by parallel electron current on electron transit timescales, faster than the Reynolds-stress-driven energy transfer rate, so the ratio $E_n/E_\sim$ stays relatively constant. While most previous analyses of the L-H transition have neglected $E_n$,[20] recent work on NSTX found $E_n > E_\sim$,[31] so we include it in our analysis.

To apply this model to the BES data from our experiment, we slightly modify Eqs. (6)–(9). To calculate $E_n$ we use the BES-estimated $\tilde{n}_{\text{rms}}/\bar{n}$ values. Since BES only makes measurements at the outboard midplane we cannot calculate flux surface averages and instead use a combination of temporal averaging and poloidal averaging over the extent of the BES array. The mean flow energy is calculated using a 1 ms centered moving average while a 100 μs centered moving average is used for fluctuating quantities, the same as for the quantities shown in Fig. 12. Fixed values of $n_0 = 1\times 10^{19}$ m$^{-3}$ and $T_{e0} = 50$ eV are used for all three MP cases.

Time traces of $E_-$, $E_n$, $E_\sim$, and $P$ at $\rho = 0.96$ leading up to and across L-H transitions with different applied MPs are shown in Fig. 13. The mean flow energy is largest in the axisymmetric case and is substantially reduced by RMPs [Fig. 13(a)], consistent with the observation in Sec. III C that RMPs reduce the L-mode $\bar{v}_\theta$ well depth. Moreover, in the axisymmetric case $E_-$ begins rising about 2 ms before the L-H transition, implying that energy is being transferred into mean shear flows. This rise does not occur when MPs are applied. We note that $E_-$ appears to begin dropping 500 μs before the L-H transition due to the 1 ms moving average window employed. We emphasize again that all inferred flow quantities go to zero after the L-H transition because $\tilde{n}_{\text{rms}}/\bar{n}$ becomes too low for velocimetry analysis to work. According to the predator-prey model, the true values of $E_n$, $E_\sim$, and $P$ do go to zero after the transition, while $E_-$ instead grows to large amplitude.

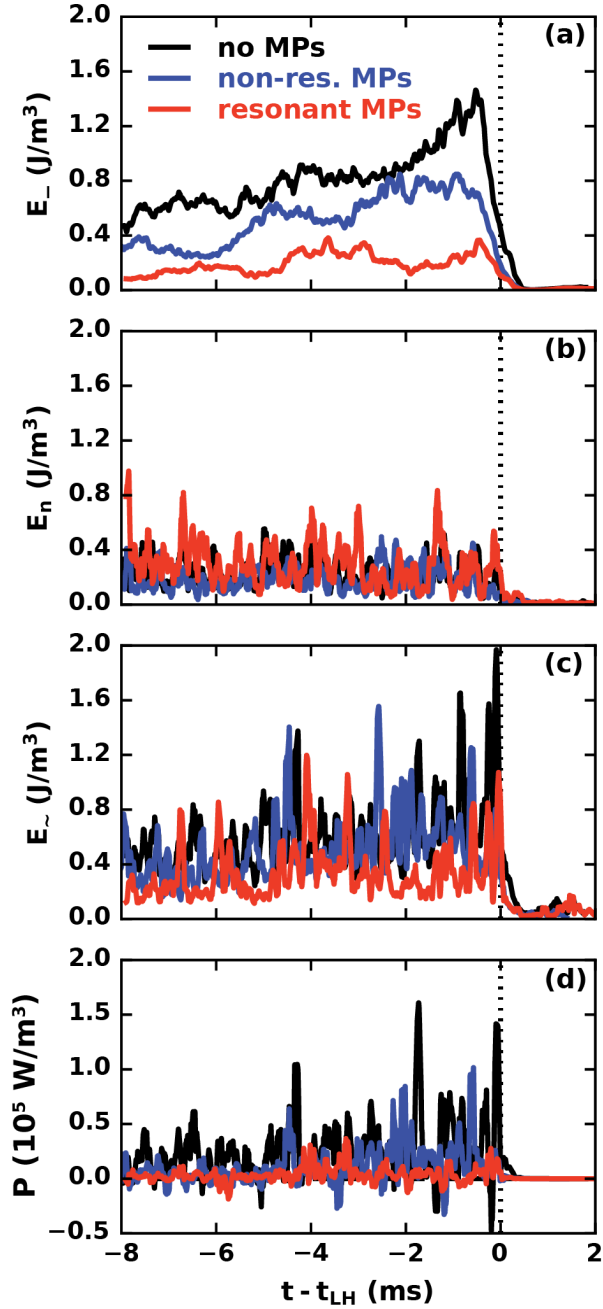

**FIG. 13.** Time traces leading up to the L-H transition at $\rho = 0.96$ (outer shear layer) of (a) mean flow energy $E_-$, (b) thermal free energy $E_n$, (c) turbulent flow energy $E_\sim$, and (d) nonlinear energy transfer rate from turbulence to flows $P$. RMPs reduce $P$ by a factor of 10, disrupting the transient turbulence suppression mechanism that is thought to trigger the L-H transition.

The thermal free energy does not change significantly leading up to the L-H transition and varies by less than 30% among the three cases [Fig. 13(b)]. It is also smaller than $E_-$ for the axisymmetric and NRMP cases, but comparable to $E_-$ in the RMP case [note that Figs. 12(a)–(c) have the same y-axis limits]. The turbulent flow energy exhibits large bursts leading up to the L-H transition that are reduced in amplitude when RMPs are applied [Fig. 13(c)]. Similarly, the nonlinear energy transfer rate shows large bursts leading up to the L-H transition that are degraded by RMPs [Fig. 13(d)]. These bursts are highly skewed in the positive direction, implying the energy transfer direction is out of turbulence and into mean flow.

To quantitatively investigate the energy dynamics, we average the quantities in Fig. 13 over a 2 ms window preceding the L-H transition. The results are shown in Table II. There are systematic uncertainties arising from simplifications made in the predator-prey model and the use of a local poloidal average instead of a flux surface average, so the absolute numerical values are only order-of-magnitude correct. Rather, what is important are the relative values between the three cases with different applied MPs. Random uncertainties are estimated using the standard error of the poloidal average and linear uncertainty propagation. The uncertainty values listed in Table II are 1.96 times the standard error, so that the mean value ± the uncertainty forms a 95% confidence interval. The relative uncertainties are larger for these quantities, generally 30–70%, than for the stationary L-mode results in Sec. III because the time averaging windows are much shorter.

**TABLE II.** Energy partition preceding L-H transitions with different applied MPs.

| MP type | None | Resonant | Non-resonant |
|---|---|---|---|
| $E_-$ (J/m$^3$) | 1.1 ± 0.4 | 0.21 ± 0.10 | 0.68 ± 0.34 |
| $E_n$ (J/m$^3$) | 0.32 ± 0.10 | 0.37 ± 0.12 | 0.27 ± 0.10 |
| $E_\sim$ (J/m$^3$) | 0.84 ± 0.27 | 0.38 ± 0.16 | 0.60 ± 0.23 |
| $P$ (10$^5$ W/m$^3$) | 0.38 ± 0.19 | 0.037 ± 0.017 | 0.20 ± 0.13 |
| $E_\sim/E_n$ | 2.6 ± 1.2 | 1.0 ± 0.6 | 2.2 ± 1.2 |
| $E_-/(E_n + E_\sim)$ | 0.93 ± 0.44 | 0.29 ± 0.16 | 0.77 ± 0.44 |
| $(E_n + E_\sim)/P$ (μs) | 30 ± 17 | 200 ± 110 | 44 ± 31 |

The ratio of turbulent flow energy to thermal free energy $E_\sim/E_n$ is near or above unity in all three cases, confirming that both $E_\sim$ and $E_n$ must be included in the analysis. Note that this is contrary to a prediction in Ref. 81 that $E_\sim/E_n \sim k_\perp^2 \rho_s^2 \ll 1$. According to the predator-prey model, the energy in mean flows must be on the same order or larger than the total turbulent energy for turbulence suppression to occur, i.e., $E_-/(E_n + E_\sim) \gtrsim 1$. The data in the axisymmetric and NRMP cases may be consistent with this criterion, but in the RMP case $E_-/(E_n + E_\sim) \ll 1$ and definitively not consistent with this criterion. RMPs therefore disrupt transient turbulence suppression on the timescale of the L-H transition.

Further evidence that RMPs disrupt transient turbulence suppression comes from estimating how fast the L-H transition is expected to occur given the total amount of turbulence energy and the nonlinear energy transfer rate. The total turbulence energy $E_n + E_\sim$ goes to zero after the L-H transition, so the reduction in turbulence energy across the transition is well-approximated by the pre-LH value in Table II. The expected timescale of the L-H transition, i.e., the time to deplete the total turbulence energy, is then given by

$$\tau_{\text{LH}} = \frac{E_n + E_\sim}{P}, \quad (10)$$

which is listed in Table II. When RMPs are applied, the value $\tau_{\text{LH}} = 200 \pm 110$ μs is about five times larger than in the axisymmetric and NRMP cases (30 ± 17 μs and 44 ± 31 μs, respectively), primarily because $P$ goes down by a factor of ten. This increase in $\tau_{\text{LH}}$ with RMPs is inconsistent with the observation that the turbulence energy drops to zero within 100 μs across the L-H transition for all three cases (Fig. 13).

An important implication of this result is that when RMPs are applied the L-H transition is *not* triggered by direct depletion of turbulence energy via Reynolds-stress-driven energy transfer. We showed in Sec. III that RMPs disrupt turbulence shear suppression in the stationary L-mode state preceding the L-H transition, meaning more *transient* turbulence suppression on the ~100 μs timescale of the transition is needed to trigger H-mode. Therefore, the predator-prey model predicts that the transient nonlinear energy transfer rate must be *larger* with RMPs. However, we instead observe it is an order of magnitude *smaller* with RMPs. This means some alternate mechanism that does not involve Reynolds stress is responsible for triggering the L-H transition when RMPs are applied. One possible mechanism is neoclassical ion orbit loss, which was shown by a recent gyrokinetic simulation of an L-H transition to play a role in concert with Reynolds stress in triggering the transition.[82] RMPs with an amplitude large enough to suppress ELMs effectively turn off the Reynolds stress mechanism that triggers the L-H transition in the axisymmetric case, which may contribute to the power threshold increase.

## V. SUMMARY

We investigated experimentally how RMPs and NRMPs alter turbulence and flows at the edge of ITER-relevant DIII-D plasmas in order to understand how RMPs raise the L-H power threshold. Our investigation was split into two parts: (i) how MPs affect turbulence-flow characteristics in the stationary L-mode state preceding the L-H transition, and (ii) how MPs affect turbulence-flow dynamics during the ~100 μs timescale of the L-H transition.

In the stationary L-mode phase, RMPs simultaneously reduce flow shear rates and raise turbulence decorrelation rates, disrupting turbulence suppression by flow shear. The flow shear rates are reduced in part because RMPs reduce the Reynolds stress drive for poloidal flow. In the fully developed H-mode state, turbulence is suppressed by flow shear and the turbulence suppression parameter $\omega_{shear}/\Delta\omega_D$ is expected to be $\geq 1$. In the L-mode state just below $P_{LH}$ without applied MPs, $\omega_{shear}/\Delta\omega_D \approx 1$, so the turbulence suppression condition is already satisfied. A small amount of *transient* turbulence suppression is then sufficient to trigger the L-H transition. RMPs reduce $\omega_{shear}/\Delta\omega_D$ to a value significantly below 1, so significantly more transient turbulence suppression is needed to trigger the L-H transition. This requires more heating power, explaining how RMPs raise $P_{LH}$. In contrast, NRMPs have no effect on turbulence decorrelation rates and only slightly reduce flow shear rates, so $\omega_{shear}/\Delta\omega_D$ decreases slightly and $P_{LH}$ is expected to increase slightly.

On the ~100 μs timescale of the L-H transition, RMPs reduce the rate of Reynolds-stress-driven energy transfer from turbulence to flows by an order of magnitude compared to axisymmetric plasmas. This implies that RMPs *reduce* transient turbulence suppression by the energy transfer mechanism. Since the results from the stationary L-mode phase show *more* transient turbulence suppression is needed to trigger the L-H transition with RMPs, we conclude that depletion of turbulence energy does not trigger the L-H transition when RMPs are applied, challenging the predator-prey model. In axisymmetric plasmas turbulence energy depletion *does* appear to trigger the L-H transition, consistent with previous work.[26]

The observation that NRMPs only weakly modify turbulence and flows and only slightly raise $P_{LH}$ motivates future experiments with mixed resonant and non-resonant MPs. This may lead to a means for mitigating the $P_{LH}$ increase, particularly if it is found that the resonance window for ELM suppression is larger than the window with significant $P_{LH}$ increase. Such experiments could also give insight into how intrinsic error fields, which contain both resonant and non-resonant components, raise $P_{LH}$, a topic of importance to ITER.[16]

**ACKNOWLEDGEMENTS**

The authors would like to thank J. D. Callen, P. W. Terry, M. J. Pueschel, K. H. Burrell, M. Leconte, and T. Stoltzfus-Dueck for helpful discussions. Part of the data analysis was performed using the OMFIT integrated modeling framework.[83] This material is based upon work supported by the U.S. Department of Energy, Office of Science, Office of Fusion Energy Sciences, using the DIII-D National Fusion Facility, a DOE Office of Science user facility, under Awards DE-FC02-04ER54698, DE-FG02-08ER54999, and DE-SC0001288. The data that support the findings of this study are available from the corresponding author upon reasonable request.